\documentclass[aps,amsfonts,pra,twocolumn,showpacs]{revtex4-1}
\usepackage{epsfig,amsmath,amssymb,bm,epsf,graphicx,psfrag}
\usepackage[all]{xy}
\usepackage{color}

\def\ket#1{\vert#1\rangle}
\def\ketbra#1{\vert#1\rangle\langle#1\vert}

\def\Longarrow{\protect\@lra}
\def\@lra{\relbar\joinrel\relbar\joinrel\relbar\joinrel%
          \relbar\joinrel\rightarrow}

\newcommand{\bc}{\begin{center}}
\newcommand{\ec}{\end{center}}
\newcommand{\be}{\begin{equation}}
\newcommand{\ee}{\end{equation}}
\newcommand{\bea}{\begin{eqnarray}}
\newcommand{\eea}{\end{eqnarray}}

\newcommand{\ncd}{\newcommand}
\ncd{\QCcns}{$QC_{\cal{C}}$}
\ncd{\QCc}{$QC_{\cal{C}}\;$}

\definecolor{libl}{cmyk}{0.2,0.1,0,0}

\begin{document}

\title{Hybrid  valence-bond states for universal quantum computation}

\author{Tzu-Chieh Wei}
\affiliation{C. N. Yang Institute for Theoretical Physics and
Department of Physics and Astronomy, State University of New York at
Stony Brook, Stony Brook, NY 11794-3840, USA}
\author{Poya Haghnegahdar}
\affiliation{Department of Physics and Astronomy, University of
British Columbia, Vancouver, British Columbia, V6T 1Z1, Canada}
\author{Robert Raussendorf}
\affiliation{Department of Physics and Astronomy, University of
British Columbia, Vancouver, British Columbia, V6T 1Z1, Canada}
\date{\today}

\begin{abstract}
The spin-3/2 Affleck-Kennedy-Lieb-Tasaki (AKLT) valence-bond state
on the hexagonal lattice was shown to be a universal resource state
for measurement-based quantum computation (MBQC). Can AKLT states of
higher spin magnitude support universal MBQC? We demonstrate that
several hybrid 2D AKLT states involving mixture of spin-2 and other
lower-spin entities, such as spin-3/2 and spin-1, are also universal
for MBQC. This significantly expands universal resource states in
the AKLT family. Even though frustration may be a hinderance to
quantum computational universality, lattices can be modified to
yield AKLT states that are universal. The family of AKLT states thus
provides a versatile playground for quantum computation.

\end{abstract}
 \maketitle

\section{Introduction}
 Quantum computation by local measurements (MBQC) was
shown to provide the same power of computation as the standard
circuit
model~\cite{Oneway,Verstraete,Gross,Oneway2,RaussendorfWei12}.
Studying various aspects of this model also offers new perspectives
and insights, such as improvement of the resource overhead in
linear-optics quantum computation~\cite{LO,Kieling}, utilization of
topological protection~\cite{TO}, renormalization and holographic
principles~\cite{Kieling,Bartlett,Miyake}, exploration of quantum
computational phases of
matter~\cite{GrossEisertEtAl,DarmawanBrennenBartlett,Murao,WeiLiKwek}
and the relation to symmetries~\cite{ElseSchwarzBartlettDoherty} and
contextuality~\cite{BrowneAnders}. There are still several aspects
in MBQC to be understood, such as complete characterization of
universal resource states and whether they can arise as unique
ground states of gapped two-body interacting
Hamiltonians~\cite{Nielsen,Chen,Cai10,ZengKwek}.

The cluster state on the square lattice was the first entangled state
found to support universal MBQC~\cite{Oneway}. Later its generalization to
graph states on many 2D regular lattices were shown to be universal~\cite{Universal}. 
Moreover, even faulty lattices can also support graph states that are universal for MBQC~\cite{Browne}. 
Given the success in graph states, one is led to ask the question: how to characterize all universal resource states?
Specifically, do there exit other families of states like graph states that provide such
versatility for quantum computational universality?
Common properties shared by members in such families can provide insight to the relation
between computational universality and  physical properties, possibly providing stepping stones towards complete
characterization of resource states, which is currently lacking. 

\begin{figure}
   \includegraphics[width=8.5cm]{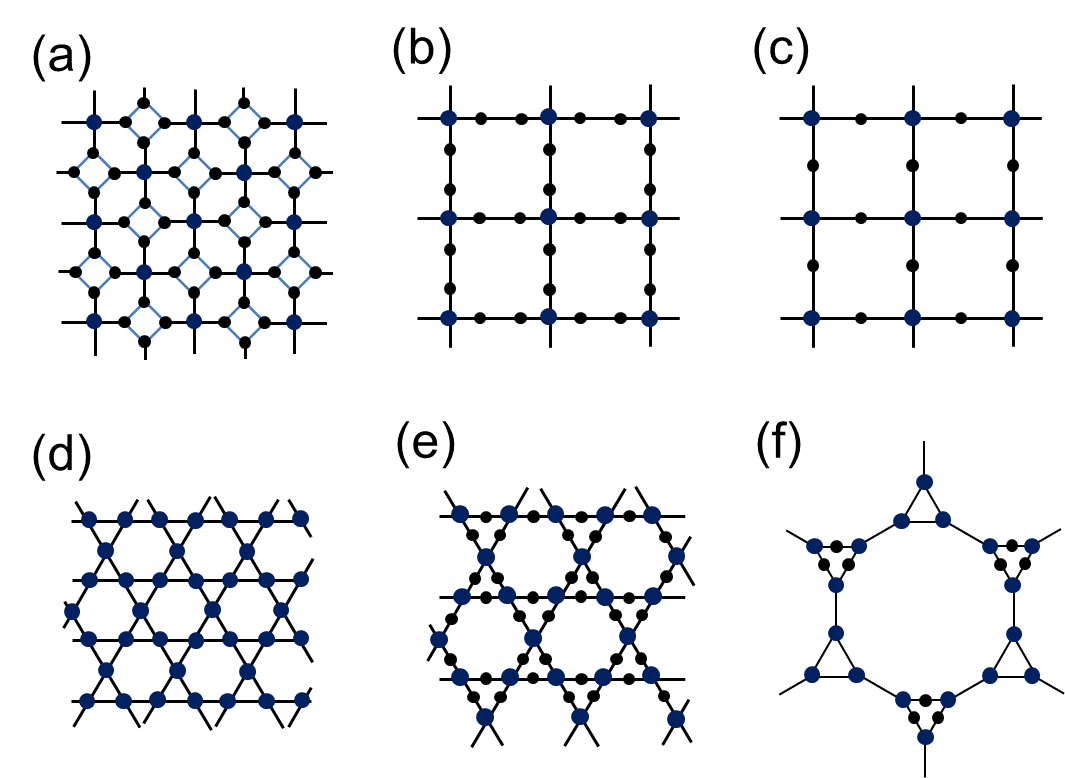}
  \caption{\label{fig:lattice} (Color online) Lattices that host various AKLT states. By construction local spin magnitude $S$ depends on
  the coordination number $z$ by $S=z/2$. (a)
   The lattice that supports a hybrid AKLT state with spin-3/2 and spin-2
   mixture. It is obtained from the square lattice by replacing in the
checker-board pattern spin-2 sites with four spin-3/2 sites.
   (b), (c) \& (e) Lattices that host spin-1 spin-2 hybrid AKLT
   states.
   (d) kagome lattice that hosts a spin-2 AKLT state.
   (f) A decorated star lattice which hosts a spin-3/2 spin-1 hybrid AKLT state.}
\end{figure}

In the following
we shall investigate the family of the Affleck-Kennedy-Lieb-Tasaki
(AKLT) states~\cite{AKLT} and significantly expand known universal
resource states in this family. Similar to graph states, AKLT states can be defined on any graph. In contrast to graph states composed uniformly of spin-1/2 entities,  
the local spin magntidue $S$ of the AKLT states depend on the number
$z$ of nearest neighboring sites, via $S=z/2$; see e.g. Fig.~\ref{fig:lattice}. The entanglement stucture of the AKLT family is therefore different from that of the graph-state family. 
Graph states do not arise as unique ground states of two-body
interacting Hamiltonians~\cite{Nielsen}, which might be a disadvantage from
the viewpoint of creating universal resource states by cooling. A finite gap in the Hamiltonian separating a unique resouce state as a ground state from excited states is thus a desirable feature~\cite{Darmawan2}.  With suitably chosen boundary conditions, AKLT states are unique ground states of certain two-body interacting Hamiltonians~\cite{AKLT}, some of which are believed to possess finite spectral gap~\cite{Ganesh,Garcia}.

The insight why AKLT states might be useful
for MBQC originated from the study of the 1D spin-1 AKLT state,
shown to be capable of simulating
 one-qubit gates~\cite{Gross,Brennen}.
But the full quantum computational universality requires higher
dimensions and was later established in the spin-3/2 AKLT state on
the honeycomb
lattice and a couple of other trivalent lattices~\cite{WeiAffleckRaussendorf11,Miyake10,WeiAffleckRaussendorf12,Wei13}.
 It is therefore natural to ask whether the university of the spin-3/2 AKLT state is fortuitous or there are other states, in particular ones with higher spins (than 3/2), that can provide universal resource for MBQC?  Furthremore, understanding why some states in the family are useful while others are not can provide connection of quantum computational universality to other physical properties.

Here we demonstrate that within the AKLT family many states defined
on various 2D lattices contain mixture of different spin entities
(such as spin $S=1/2,1,3/2$ and $2$) and provide universal resources
for MBQC. (We note that spin-1/2 entities can
occur at the boundary of the lattice.) The universality of this family is richer than the cluster
family. Lattices possessing geometric frustration (i.e., loops with
odd number of sites) may not support universal resource states but
they can be modified (or decorated) so that the associated AKLT
states are universal.  The emerging picture from our study is the
following. AKLT states involving spin-2 and other lower spin
entities are universal if they reside on a two-dimensional
(geometric) frustration-free lattice with any combination of spin-2,
spin-3/2, spin-1 and spin-1/2 (consistent with the
lattice), provided that spin-2 sites are not neighbors.
The constraint that the spin-2 states are not neighbors is due to the open question 
that whether AKLT states of uniform spin-2 entities can be universal, and
it may be lifted if the anwser to that question is affirmative. We also note that,
however, the existence of geometric frustration may not necessarily
render the states non-universal, as we shall also demonstrate.

The task of proving universality of higher-spin AKLT states poses a
greater challenge, as a straightforward extension of the generalized
measurement or positive-operator-valued-measure measurement (POVM)
for the trivalent case does not work. Additional elements must be
introduced to complete the spin-2 POVM~\cite{LiEtAl}, and as a
result  the known reduction of the AKLT states to graph states may
no longer work. We shall provide a construction of spin-2 POVM (see
Eqs.~(\ref{POVMspin2}) below) that allows us to overcome this
difficulty.
 We consider AKLT states on a few lattices, some of which are shown
 in Fig.~\ref{fig:lattice}, but the applicability of our method goes beyond these. For example, the lattice Fig.~\ref{fig:lattice}a
  can be regarded as modified from the square lattice by replacing in the
checker-board pattern spin-2 sites with four spin-3/2 sites. On this
lattice, spin-2 sites do not reside as neighbors, but are surrounded
by other spin-3/2 sites. Spin-3/2 sites can be dealt easily with a
POVM composed of three directions (x, y and z); see
Eq.~(\ref{POVMspin32})~\cite{WeiAffleckRaussendorf11}. Such
modification makes the demonstration of the universality involving
higher-spins possible via treating potential leakage errors on these
sites. In a similar manner we shall also examine two other lattices
(Figs.~\ref{fig:lattice}b and c) which support AKLT states with
mixture of spin-2 and spin-1 entities. Moreover, we shall
investigate the universality for the spin-2 AKLT state on the kagome
lattice (Fig.~\ref{fig:lattice}d) and the spin-mixture AKLT states
on the decorated kagome and star lattices (Fig.~\ref{fig:lattice}e
and f), respectively. For convenience, we shall refer to these AKLT states defined on graphs with
non-uniform vertex degrees as {\it hybrid} AKLT states or hybrid valence-bond states.

The structure of the remaining of this manuscript is as follows. In Sec.~\ref{sec:POVM} we discuss an important 
in gredient of our proof for universality, i.e., the generalized measurement. It enables projection from four- or five-level states to effective two-level states, i.e., qubits. In Sec.~\ref{sec:scenario} we describe a strategy to perform simulations without sampling the exact distribution of POVM outcomes. It is a worst-case scenario and therefore provides a `lower bound' on universality. In Sec.~\ref{sec:geometric} we discuss consequences of geometric frustration for AKLT states on kagome and star lattices and provide some `decoration' on these lattices that yied universal AKLT states. We summarize in Sec.~\ref{sec:conclude}. 
\section{The generalized measurement for reduction to qubits}
\label{sec:POVM}

One approach for universality is the so-called quantum state
reduction~\cite{Chen10}, i.e., to show that the state in question
can be converted, via local measurements, to a known resource state,
in particular some 2D graph state.
 The POVM for spin-3/2 sites consists of three
rank-two elements~\cite{WeiAffleckRaussendorf11}
\begin{equation}
\label{POVMspin32}
F_\alpha=\sqrt{\frac{2}{3}}\,(\ketbra{{S_\alpha=\frac{3}{2}}}\,+\,\ketbra{{S_\alpha=\frac{-3}{2}}}),
\end{equation}
with $\alpha=x,y,z$. The POVM for the spin-2 sites consists of three
rank-two elements and three additional rank-one elements:
\begin{subequations}
\label{POVMspin2}
\begin{eqnarray}
\!\!\!\!\!F_\alpha&=&\sqrt{\frac{2}{3}}\,(\ketbra{{S_\alpha\!=\!2}}+\ketbra{{S_\alpha\!=\!-2}})\\
\!\!\!\!\!K_\alpha&=&\sqrt{\frac{1}{3}}\,\ketbra{\phi_\alpha^-},
\end{eqnarray}
\end{subequations} where
 $\alpha=x,y,z$ and
$\ket{\phi_\alpha^\pm}=(\ket{{S_\alpha\!=\!2}}\pm\ket{{S_\alpha\!=\!-2}})/\sqrt{2}$.
It can be verified that the completeness relation is satisfied:
$\sum_\alpha {F}^\dagger_\alpha {F}_\alpha +\sum_\alpha
{K}^\dagger_\alpha {K}_\alpha=\openone$.

It has previously been
shown~\cite{WeiAffleckRaussendorf11,WeiAffleckRaussendorf12} that
for the state $|\psi\rangle\sim\otimes_{v\in V}
{F}_{\alpha_v,v}|\psi_{\rm AKLT}\rangle$ (where $v$ is a site index)
is an encoded graph state on domains, regardless of
$\alpha_v$'s (the proof applies to any spin $S$). 
By a {\it domain\/} we mean a collection of connected sites on the original
lattice that have outcomes associated with the $F$'s with the same labeling
$\alpha$. A domain is effectively a qubit.
Outcomes
associated with $K_\alpha$ are thus undesired, and we might need to regard
them as ``errors''~\cite{referee}. Nevertheless, the ${K}$ operators can be
rewritten as $ {K}_\alpha= \sqrt{{1}/{2}}
 \ketbra{\phi_\alpha^-}{F}_\alpha$. This suggests that we
can regard the outcome associated with ${K}_\alpha$ as arising from
a two-step process: (1) first a result in the outcome associated
with ${F}_\alpha$ is obtained; (2) then a further measurement is
done in the basis $\ket{\phi_\alpha^\pm}$ and the result
$\ket{\phi_\alpha^{-}}$ is obtained.  A measurement in the basis
$\ket{{\phi}_\alpha^{\pm}}$ corresponds to a measurement in the
effective logical $X\equiv\sigma_x$ or $Y\equiv\sigma_y$
basis~\cite{Note2}, respectively.

\begin{figure}
   \includegraphics[width=6cm]{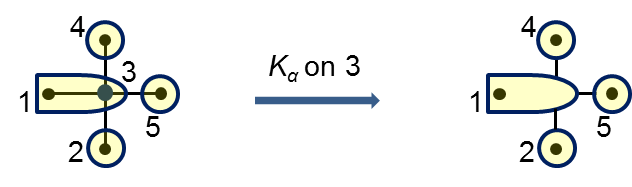}
  \caption{\label{fig:codereduction}  (Color online) Code reduction.  Sites 1 \& 3 are in the same domain, and 2,4,5 are each a distinct domain. A POVM $K_\alpha$ on the spin-2 site 3 is equivalent to
  removing site 3 without affecting the graph. Shapes such as circles (single-site) and half-ovals (multi-site) are domains. On the l.h.s. an internal edge (between sites 1 and 3) is shown.}
\end{figure}

By construction the $K_\alpha$'s outcomes only occur on spin-2
sites. For each spin-2 site that is contained in a {\it
multi-site\/} domain, the effect of the measurement in the basis
$\ket{{\phi}_\alpha^\pm}$ correspond to {\it code reduction\/},
namely, shrinking the number of sites inside the domain by one without affecting the
quantum correlations of the present domain with others (i.e., the
graph remains the same)~\cite{WeiAffleckRaussendorf11}; see e.g.
Fig.~\ref{fig:codereduction}. Therefore, it  has no effect on the
graph for the encoded graph state. However, for a {\it
single-site\/} spin-2 domain, the measurement outcome ${K}_\alpha$
amounts to either a logical $X$ or $Y$ measurement on this logical
qubit of the graph state. For the purpose of discussion we refer to
them as X or Y {\it undesired measurement\/}, respectively. Pauli
measurements on a graph state simply results in another graph state,
whose graph can be easily deduced from simple rules~\cite{Hein}.
However, these Pauli measurements do not preserve planarity 
of graphs. In order to apply the percolation argument for planar
graphs~\cite{WeiAffleckRaussendorf11,WeiAffleckRaussendorf12}, we need to somehow actively 
perform further measurement to recover planarity of the graphs, as will be discussed below.

\begin{figure}
   \includegraphics[width=8cm]{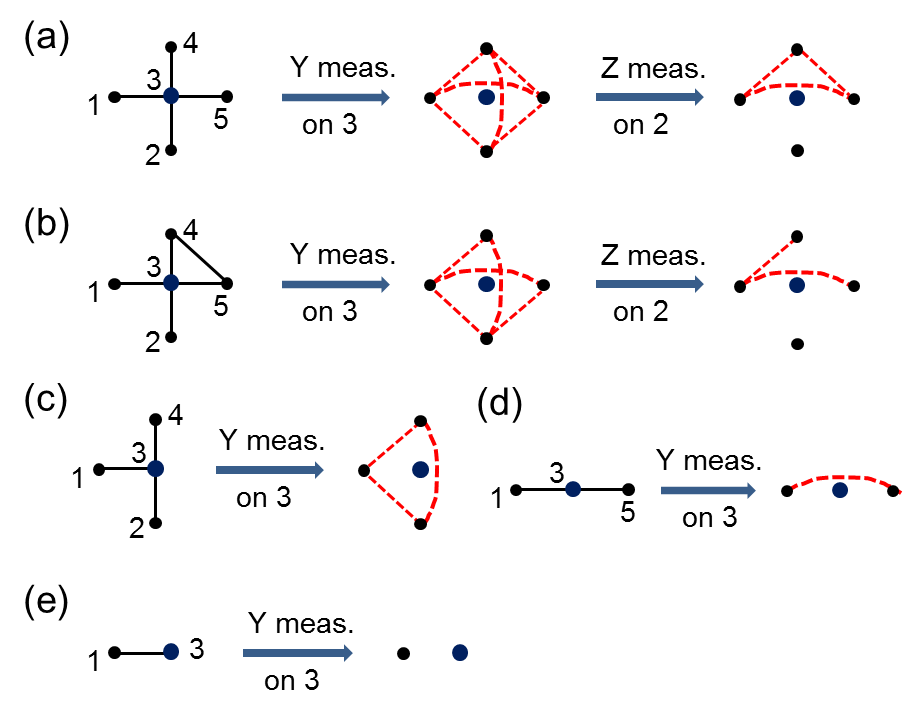}
  \caption{\label{fig:graphrules}  (Color online) Graph transformation rules by Y  measurement on a
  single-site spin-2 domain (site 3). For simplicity we assume every site is a domain. (a) and (b) illustrate the case where four distinct domains are connected to
  center spin-2 site. In (b)  there is an additional edge between domains 4 and 5.
  To further make the graph remain planar, a Pauli Z measurement is made on any of the neighboring domain, say, 2.
  (c) shows the case where three distinct domains are connected to the spin-2 site.
  (This case can arise, e.g., as
  one of the neighboring domains was deleted in  the second step of (a) or (b) associated with
  other spin-2 site.)  (d) and (e) exemplify the cases of, respectively, two and one domain connected
  to a spin-2 site. Note that the above list does not exhaust all possibilities
   but just serves to illustrate that the undesired Y measurement can be
  treated to maintain planarity.
}
\end{figure}

The above discussions suggest that one approach for establishing
universality is to employ certain procedure to recover the planarity
of the graphs, caused by the ${K}_\alpha$ outcomes on single-site
spin-2 domains.  Consider the undesired Y measurement. Its effect is
to induce local complementation on the graph~\cite{Hein}. When the
spin-2 domain is connected to three or few other domains, the
planarity is preserved. When it is connected to four other domains
(four is the maximum due to the geometry considered here),  we can
simply apply an additional Pauli Z measurement on any neighboring
domain so as to recover planarity. See Fig.~\ref{fig:graphrules} for
illustration of these measurements. The undesired X measurements are
more troublesome, however, as they can result in longer-range
connectivity (edges), and the planarity is more complicated to
restore (it also depends on which reference neighbor to use for
applying the graph rule~\cite{Hein}). We shall take the simplest
approach  by measuring logical Z on all their neighbors, thereby
removing both the domains with undesired X measurement and their
neighboring domains. This can be costly, but simplifies the
simulations. In summary, we shall first treat all the undesired X
measurement and then Y measurement to recover the full planarity.

\begin{figure}[t]
 \begin{tabular}{l}
    (a)\\
\vspace{-1cm} {\includegraphics[width=.5\textwidth]{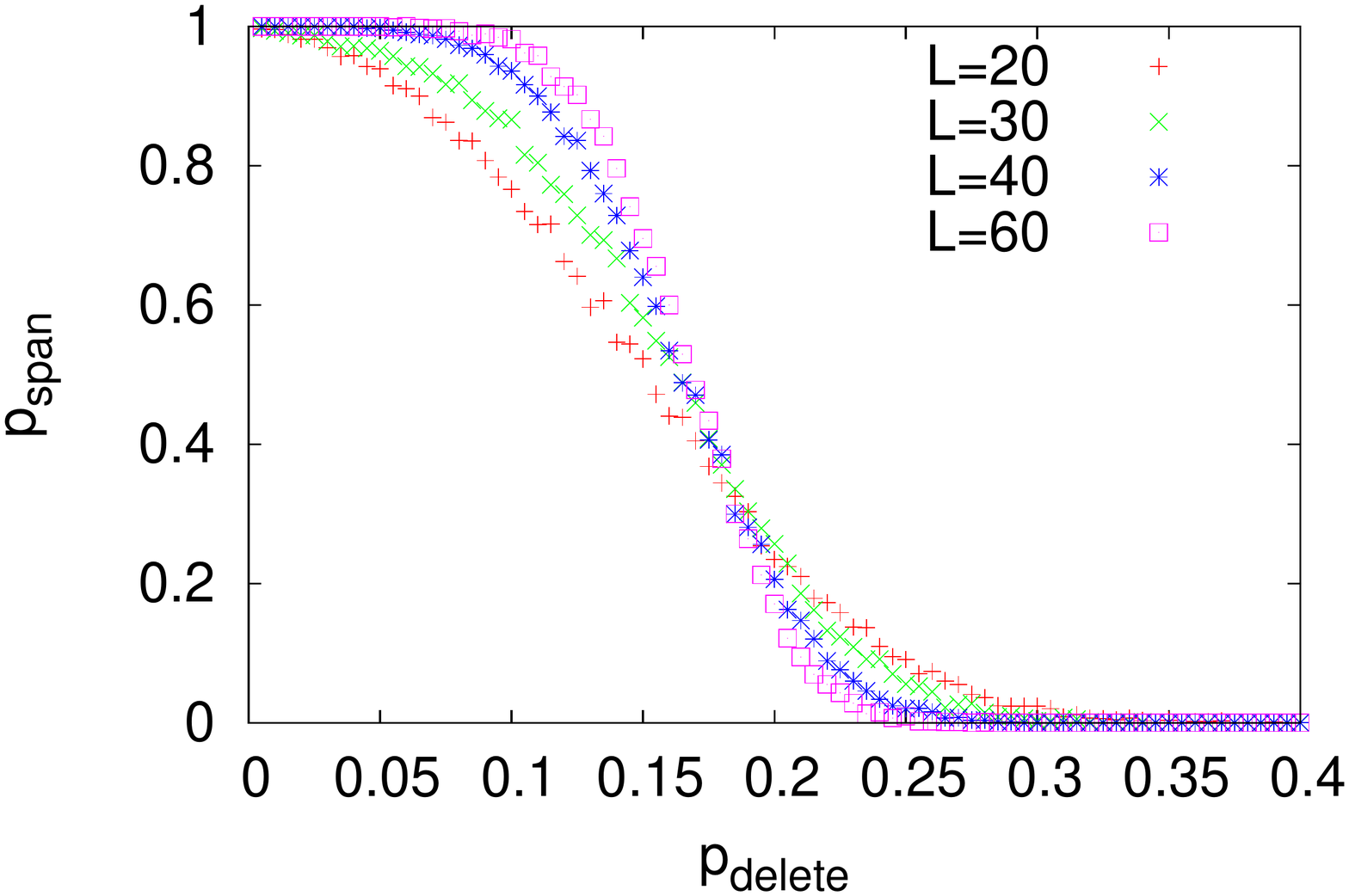}}
 \\
 (b)\\
 \vspace{-1cm} {\includegraphics[width=.5\textwidth]{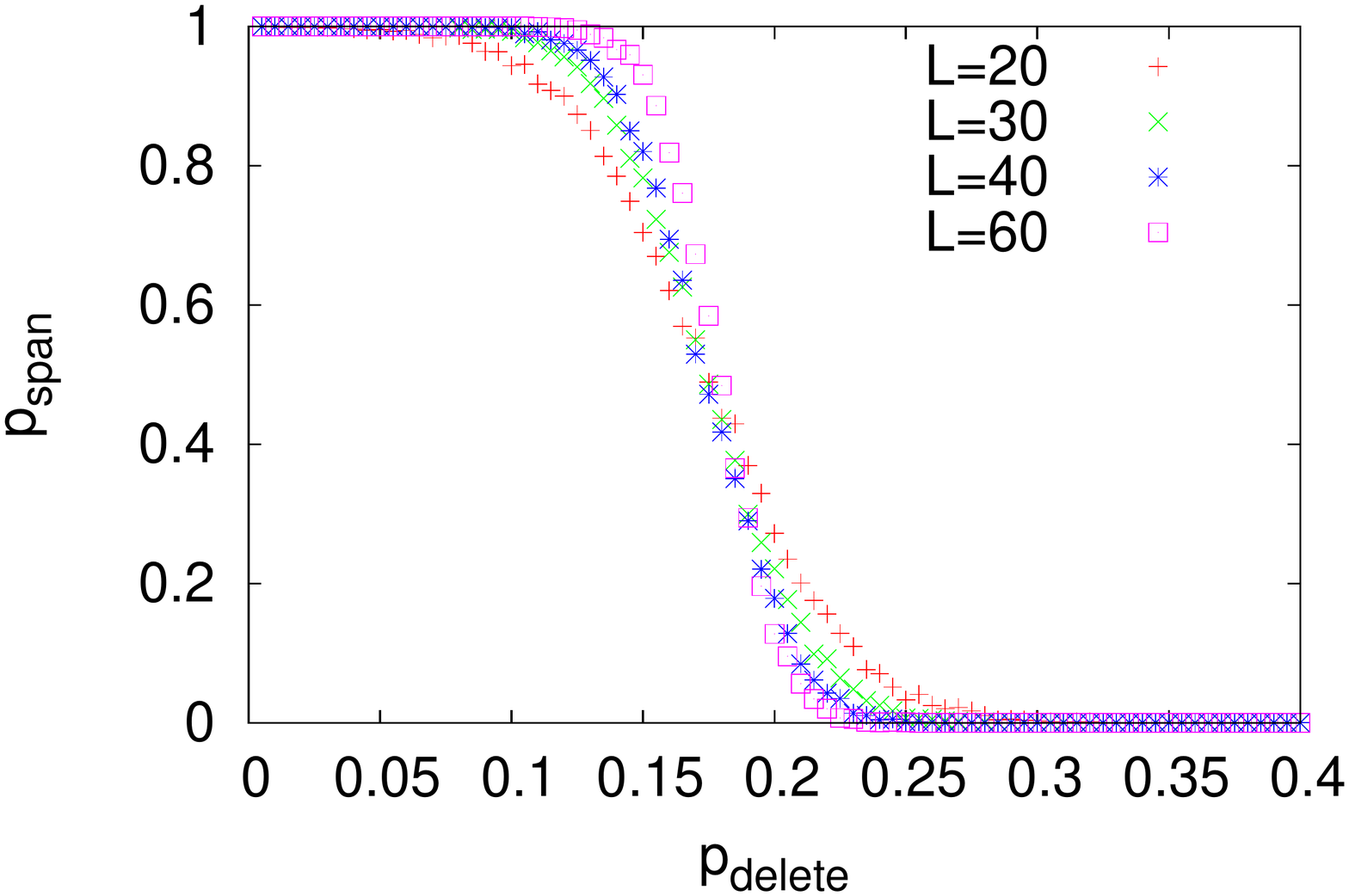}}
 \\
  (c)\\
 \vspace{-1cm} {\includegraphics[width=.5\textwidth]{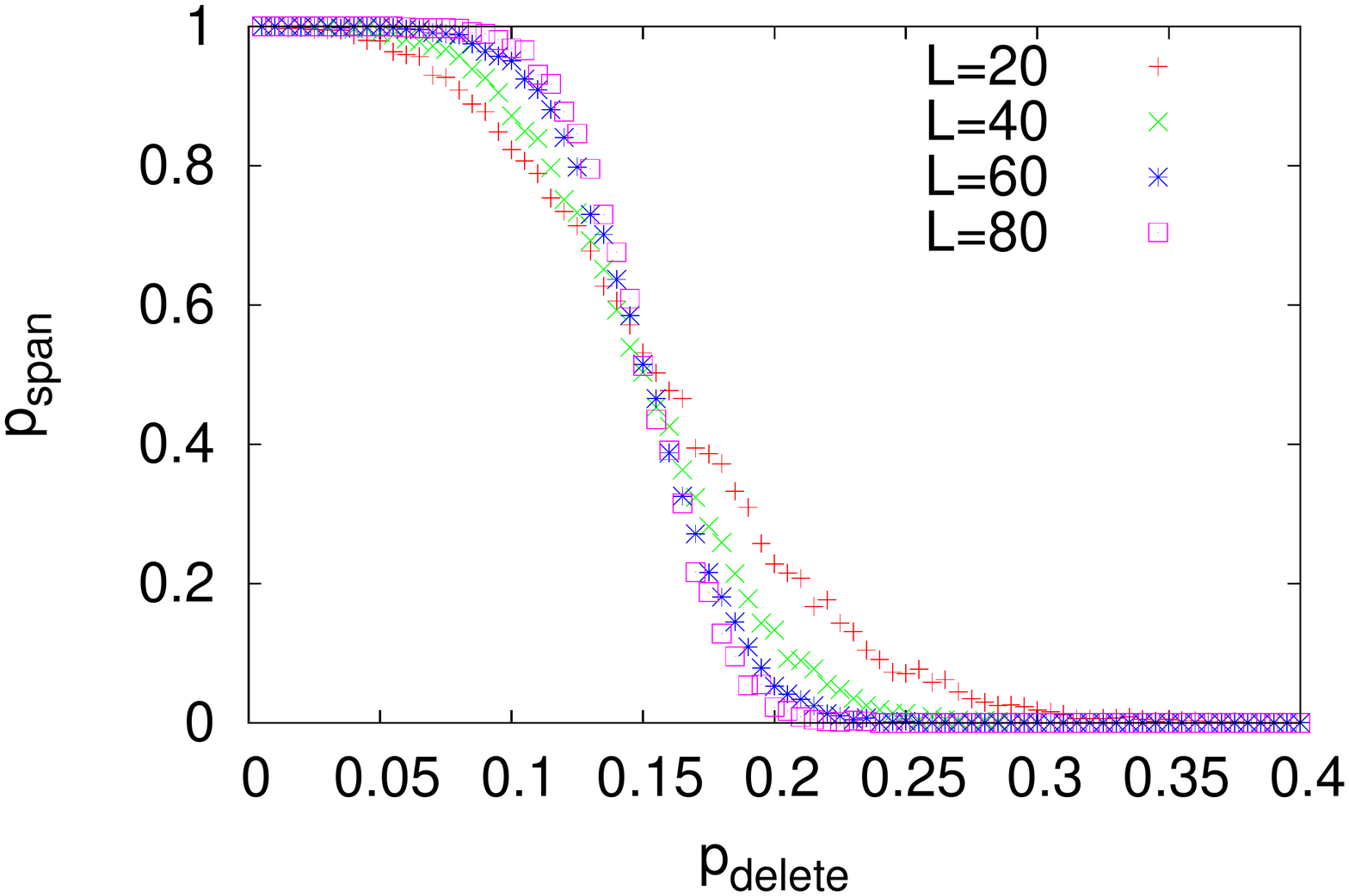}}
  \end{tabular}
        \caption{(Color online)  Site percolation study on
         the random graphs of domains resulting from the worst-case scenario procedure on
         the AKLT states on the lattices of (a) Fig.~\ref{fig:lattice}a with a total number of
         quantum spins (or sites) $N=5L^2$,  (b) Fig.~\ref{fig:lattice}b with $N=5L^2$, and (c) Fig.~\ref{fig:lattice}c with $N=3L^2$. Note that the quantities in both axes are dimensionless.
In all cases, there is a phase transition (as measured by the
probability of a spanning path $p_{\rm span}$) from supercritical to
subcritical phase as the probability of deletion ($p_{\rm delete}$)
increases. The estimated thresholds of $p_{\rm delete}$ are:
$0.18(1)$, $0.19(1)$, and $0.11(1)$, respectively. Nonzero deletion
thresholds show that the resulting random graphs after the procedure
(before site deletion) are deep in the supercritical phase. }
\label{fig:perco}
\end{figure}
\section{Worst-case scenario}
\label{sec:scenario} We have just described
a procedure to deal with $K_\alpha$ outcomes. However, in order to
perform simulations to determine the computational universality, we
need a method to sample ${F}$'s and ${K}$'s from the exact
distribution according to the correlation in the AKLT states. Via
exact sampling, we can also determine the probability of success for
reducing the AKLT to a universal graph state.
 The  exact sampling amounts to knowing the probability
 distribution of
measuring Pauli X/Y on some subset of qubits in a graph state.
Although it is a linear-algebra problem in principle, obtaining the
solutions and performing the sampling may be nontrivial.
Fortunately, it turns out that we can avoid the exact sampling by
considering the `worst scenario' and still demonstrate universality
(with unit probability) for various AKLT states.

Using the 2-step viewpoint of ${K}$'s, we first sample POVM outcomes
according to ${F}$'s, which was solved in
Ref.~\cite{WeiAffleckRaussendorf11}. The exact sampling could be
obtained by assigning some spin-2 sites to flip from ${F}_\alpha$ to
${K}_\alpha$ according the exact probability distribution had one
solved the linear-algebra problem mentioned above~\cite{Note3}.
Instead of sampling the exact distribution we then use
 what we call `the worst-case scenario': {\it all\/} the
 spin-2 sites with ${F}_\alpha$ are flipped to ${K}_\alpha$. Then
we implement the procedure in the previous section to recover full
planarity. The resulting graphs are checked to see if they reside in
the supercritical phase of percolation or not. Certifying that they
are deep in the phase can be done by performing percolation
`experiment' (such as deleting vertices or edges) on these
graphs~\cite{WeiAffleckRaussendorf11}. If the graphs (before
deletion) are in the supercritical phase, as the probability of
deletion increases, there will be a clear signature of phase
transition, beyond which no spanning cluster exists. 

In order for universal quantum computation on graph states, the graph needs to possess a macroscopic number of traversing paths. This means that that the random graphs generated by our procedure need to reside inside the supercritical phase. Sitting right at the percolation transition is not sufficient for universal MBQC.  We have checked the for the size $L$ large enough, $p_{\rm span}$ is unity for random graphs resulting from the procedure on the AKLT states in Fig.~\ref{fig:lattice}a-c. But checking just this is not sufficient for establishing universality, as this does not rule out that the possibility that the graphs are sitting at the transition point nor that the graphs may be close to one-dimensional-like structure or star graphs (both being non-universal for MBQC).  We have
further performed site percolation simulation
on the resultant graphs, shown in Fig.~\ref{fig:perco}a-c. It is
seen that the graphs after the above procedure reside in the
supercritical phase (with unit probability), even in the worst-case
scenario. 
The significance of performing  $p_{\rm span}$ vs. $p_{\rm delete}$ simulations is therefore to first prove an existence of phase transition, justifying the separation of two distinct phases and therefore the existence of macroscopic numbers of traversing paths (in particular at $p_{\rm delete}=0$) in the supercritical phase of percolation. This shows that the  hybrid AKLT states on Fig.~\ref{fig:lattice}a-c are universal resources for MBQC.

\section{Geometric frustration: kagome and star lattices}
\label{sec:geometric}
For antiferromagnets, geometric frustration often gives rise to unexpected features. It was previously found in the star lattice that geometric frustration  may inhibit the AKLT state from being universal~\cite{Wei13}.  Here we argue that it is the case in the kagome lattice, but show that for both lattices, we can decorate them so as to obtain universal AKLT states.
\subsection {The spin-2 AKLT state on the kagome lattice} For
the spin-2 only AKLT states, the worst-case scenario cannot be
directly applied, as all the sites in each domain have some
probability to be flipped from ${F}_\alpha$ to ${K}_\alpha$.
However, even if we assume the {\it best-case scenario\/}, i.e., without
flipping ${F}_\alpha$'s to ${K}_\alpha$'s, the graphs from sampling
only the ${F}$'s outcomes do not possess traversing paths for large
enough lattice size.  Thus, the spin-2 AKLT state on the kagome is
not likely universal. We remark that what we have not ruled out is the possibility of other POVMs that
might have enabled universal quantum computation.

The main reason for the lack of universality is due to the
antiferromagnetic property in the AKLT state and the geometric
frustration in the lattice. Due to the frustration in a triangle,
the POVM outcomes with $(x,x,x)$, $(y,y,y)$ and $(z,z,z)$ cannot
appear, similar to the star lattice~\cite{Wei13}. This gives rise to
an average probability of $p_{\rm delete}^{\rm [bond]}= 1/2$ to
remove an edge from an triangle, or equivalently the probability of
occupying an edge being $1-p_{\rm delete}^{\rm [bond]}=1/2$, lower
than the bond percolation threshold for the kagome lattice $p_{\rm
th}^{\rm [bond]}\approx 0.5244$. Therefore, the random graphs are
not in the supercritical phase, implying non-universality for the
original AKLT state. This intuitive argument does not take into
account correlated errors and hence correlated edge deletion, but that is
exactly what our simulations have dealt with.

\begin{figure}[t]
 \begin{tabular}{l}
    (a)\\
\vspace{-1cm} {\includegraphics[width=0.5\textwidth]{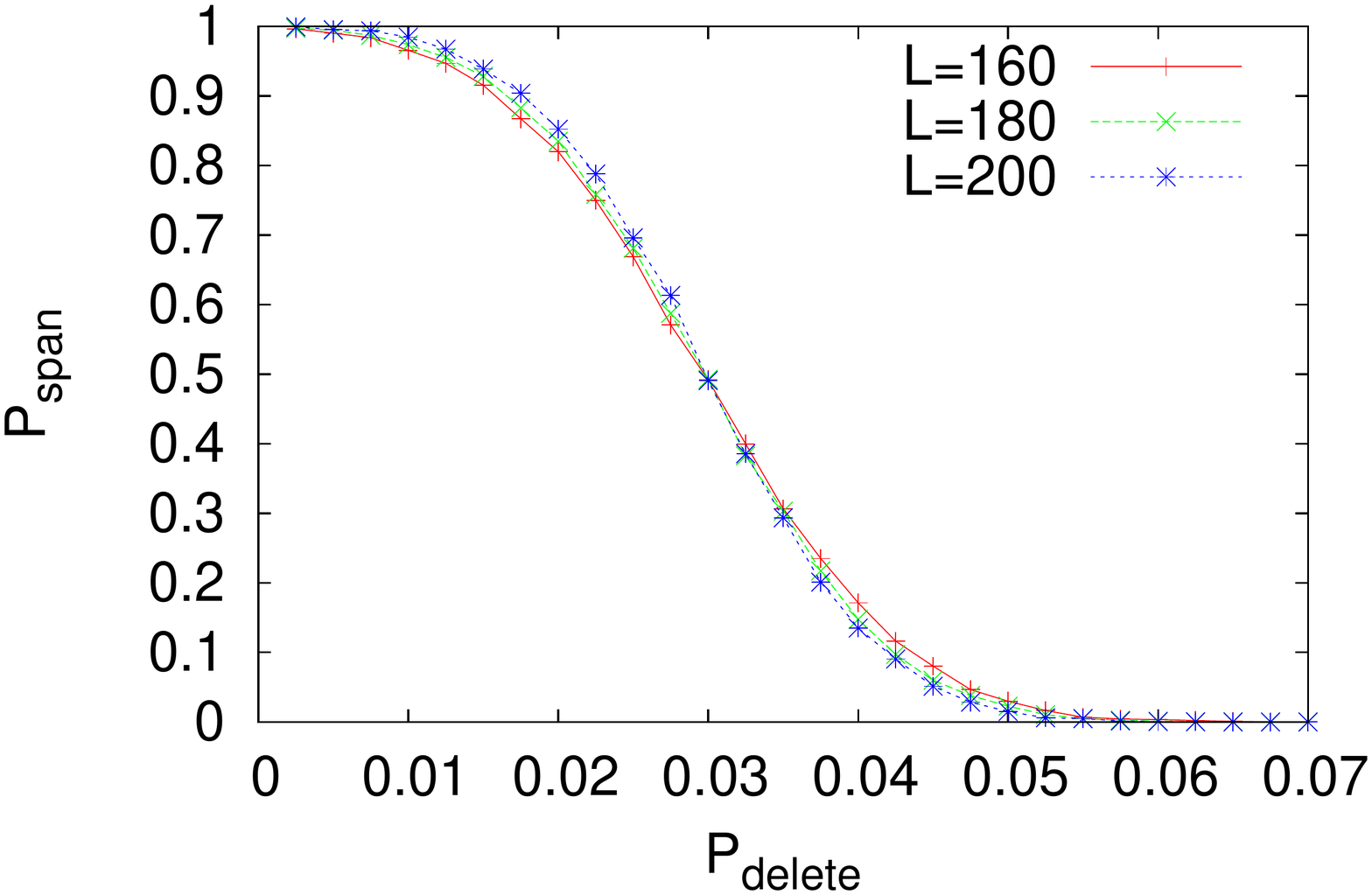}}
 \\
 (b)\\
 \vspace{-1cm} {\includegraphics[width=0.5\textwidth]{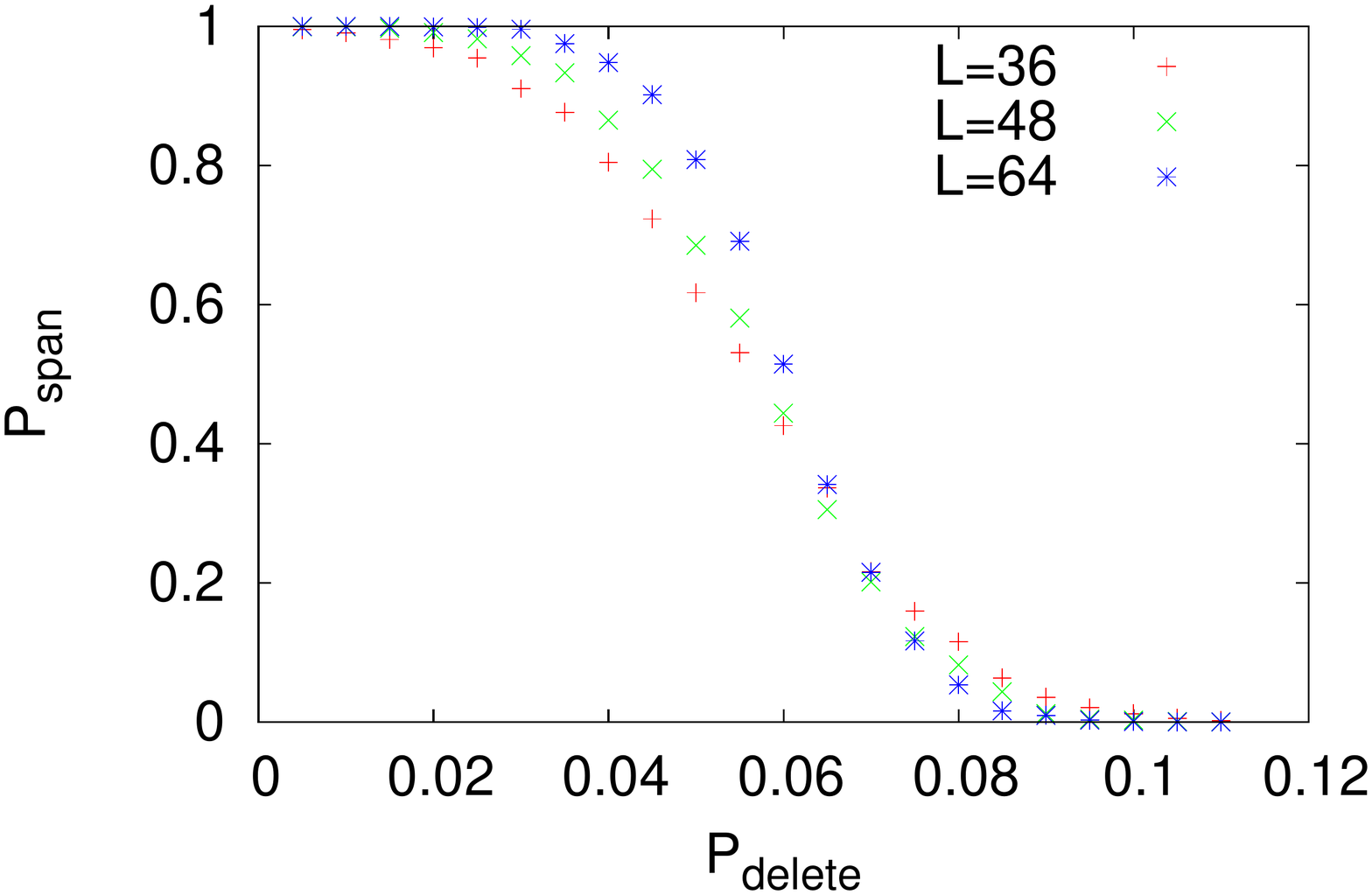}}
  \end{tabular}
        \caption{(Color online)  Site percolation study on
         the random graphs of domains resulting from the worst-case scenario procedure on
         the AKLT states on the lattices of (a) the decorated kagome, Fig.~\ref{fig:lattice}e, with a total number of
         quantum spins (or sites) $N=9L^2$ and (b) the decorated star, Fig.~\ref{fig:lattice}f, with $N=9L^2$. Note that the quantities in both axes are dimensionless.
In both cases, there is a phase transition from supercritical to
subcritical phase as the probability of deletion ($p_{\rm delete}$)
increases  (as seen from the crossing of $p_{\rm span}$ curves for different sizes). In (a) the lines are used to guide the eyes.  }
\label{fig:perco2}
\end{figure}

\subsection{ Decorated Kagome lattice: spin-1 spin-2
mixture\/} To put to test the emerged picture that the AKLT state
on any frustration-free lattice of at most degree-4 is universal, we
decorate the Kagome lattice by putting on each edge a vertex,
removing frustration. This decorated lattice then hosts a spin-1
spin-2 mixture hybrid AKLT state. For the lattice with $L\times L$ unit
cells (each having $3$ spin-2 particles and $6$ spin-1 particles per
cell) there are a total number $N=9L^2$ of spins. Our simulations
show that for $L$ large enough (e.g. $L\gtrsim 140$) our procedure
yields random graphs in the supercritical phase with almost unit
probability. The results of site percolation simulations are shown in Fig.~\ref{fig:perco2}a. Even though the threshold of $p_{\rm delete}$ is small, around 0.03, it is still nonzero and therefore the random graphs are residing inside the supercritical phase. In summary, by decorating a frustrated lattice we are able to make
the resultant AKLT state universal.

\subsection { Decorated star lattice} Whether the
existence of geometric frustration will completely destroy the
universality may depend on the particular lattice. It was shown that
the spin-3/2 AKLT state on the star lattice is not likely
universal~\cite{Wei13}. But we can decorate the star lattice by
placing an additional vertex (which corresponds to a spin $S=1$
site) on each edge of all upside-down triangles; see
Fig.~\ref{fig:lattice}f. There is {\it still\/} geometric frustration due to
the remaining triangles and one might expect the associated AKLT
state might not be universal. However, our simulations show that the
resulting graphs after POVM are with probability one in the
supercritical phase. The results of site percolation simulations are shown in Fig.~\ref{fig:perco2}a. 
Therefore, the associated spin-3/2 spin-1
hybrid AKLT state, even in presence of frustration, is still a
universal resource.
\section{Concluding remarks}
\label{sec:conclude}
We have shown that many 2D hybrid AKLT states (see Fig.~\ref{fig:lattice})
involving mixture of spin-2 and other lower-spin entities are also
universal for MBQC. The results are nontrivial as they demonstrate
that AKLT states with higher spins than 3/2 can still be universal.
We have also demonstrated that even though the spin-2 AKLT state on
the kagome lattice (Fig.~\ref{fig:lattice}d) is not likely
universal, one can decorate the lattice by adding additional sites
so that the resultant AKLT state becomes universal. Moreover, the
existence of geometric frustration does not necessarily destroy the
universality, as demonstrated in spin-3/2 spin-1 hybrid AKLT state
on the decorated star lattice. The following picture on the
universality of AKLT states emerges: any 2D frustration-free lattice
with any combination of spin-2, spin-3/2, spin-1 and spin-1/2,
except that spin-2 sites are not neighbors, will host a universal
AKLT state. To lift the constraint that the spin-2 sites are not neighbors, one would need to
resolve the question question whether the spin-2 AKLT state
on the square lattice remain universal. Furthermore, the effect of frustration may not be so severe as to
completely destroy universality. Otherwise by simple decoration to remove frustration a resulting
hybrid AKLT state can be universal. The family of 2D AKLT
states contain many members that are universal, providing a
nontrivial and versatile playground for quantum computation.

\medskip \noindent {\bf Acknowledgment.}  This work was supported by the
National Science Foundation under Grants No. PHY 1314748 and No. PHY
1333903 (T.-C.W.) and by NSERC, Cifar and PIMS (R.R.).

\end{document}